\documentclass[sigconf,nonacm]{acmart}
\usepackage[ruled,vlined,linesnumbered]{algorithm2e}
\usepackage{algorithmic}
\usepackage{amsmath}
\usepackage{graphicx}
\usepackage{subcaption}

\usepackage{caption}
\usepackage{hyperref}
\newcommand{\DateQuery}{Date Query}
\newcommand{\WeatherQuery}{Weather Query}
\newcommand{\FlightQuery}{Flight Query}
\newcommand{\TrainQuery}{QTrain Query} 
\newcommand{\PriceComp}{Price Comp.}   
\newcommand{\ScheduleQuery}{Schedule Query}
\newcommand{\PrefQuery}{Pref. Query}
\newcommand{\IntentClf}{Intent Clf.}     
\newcommand{\ParamID}{Param. ID}         
\begin{document}

\title{A Plan Reuse Mechanism for LLM-Driven Agent}

\author{Guopeng Li}
\affiliation{
  \institution{University of Science and Technology of China}
  \country{Hefei, China}
}
\author{Ruiqi Wu}
\affiliation{
  \institution{University of Science and Technology of China}
  \country{Hefei, China}
}
\author{Haisheng Tan}
\authornote{Corresponding author.}
\affiliation{
  \institution{University of Science and Technology of China}
  \country{Hefei, China}
}


\begin{abstract}
Integrating large language models (LLMs) into personal assistants, like Xiao Ai and Blue Heart V, effectively enhances their ability to interact with humans, solve complex tasks, and manage IoT devices. Such assistants are also termed LLM-driven agents. Upon receiving user requests, the LLM-driven agent generates plans using an LLM, executes these plans through various tools, and then returns the response to the user. During this process, the latency for generating a plan with an LLM can reach tens of seconds, significantly degrading user experience. Real-world dataset analysis shows that about 30\% of the requests received by LLM-driven agents are identical or similar, which allows the reuse of previously generated plans to reduce latency. However, it is difficult to accurately define the similarity between the request texts received by the LLM-driven agent through directly evaluating the original request texts. Moreover, the diverse expressions of natural language and the unstructured format of plan texts make implementing plan reuse challenging. To address these issues, we present and implement a plan reuse mechanism for LLM-driven agents called AgentReuse. AgentReuse leverages the similarities and differences among requests’ semantics and uses intent classification to evaluate the similarities between requests and enable the reuse of plans. Experimental results based on a real-world dataset demonstrate that AgentReuse achieves a 93\% effective plan reuse rate, an F1 score of 0.9718, and an accuracy of 0.9459 in evaluating request similarities, reducing latency by 93.12\% compared with baselines without using the reuse mechanism.
\end{abstract}


\maketitle
\thispagestyle{plain} 
\pagestyle{plain}   
\begingroup
\renewcommand{\thefootnote}{}
\footnote{This paper is an English version of \textit{A Plan Reuse Mechanism for LLM-Driven Agent} (\href{https://crad.ict.ac.cn/en/article/doi/10.7544/issn1000-1239.202440380}{https://crad.ict.ac.cn/en/article/doi/10.7544/issn1000-1239.202440380}), published in 2024 in the Journal of Computer Research and Development.}
\endgroup
\addtocounter{footnote}{-1} 
\section{Introduction}
\label{sec:intro}


In recent years, with the growth in the number of IoT devices and the rise of technologies such as artificial intelligence (AI) and edge computing, the Artificial Intelligence of Things (AIoT) has received widespread attention as a future trend in IoT development \cite{wang2025scalability}. As typical AIoT applications, personal assistants such as Xiao Ai \cite{xiaomi_xiaoai} and Huawei Xiaoyi \cite{huawei_voice} have improved the utility and integration of IoT devices in fields like smart homes and smart transportation by simplifying the user control process, thereby enhancing the interactivity and automation levels of AIoT systems.

Recently, Large Language Models (LLMs) such as ChatGPT, DeepSeek, and Qwen have demonstrated exceptional capabilities in understanding and generating human language, bringing new opportunities for the development of personal assistants \cite{dong2023next}. Manufacturers like Google, Xiaomi, and VIVO have launched personal assistants based on large models, aiming to leverage LLMs to improve abilities in interacting with humans, completing complex tasks, and efficiently managing and controlling applications and various IoT devices \cite{li2024personal}. Such assistants are termed \textit{personal LLM-driven agents} \cite{weng2023llm}, or simply \textit{agents} in this paper.

An LLM-driven agent uses an LLM as its core controller, supplemented by capabilities such as planning, memory, and tool use \cite{wang2024survey}. By coordinating these different capabilities, it completes tasks submitted by users. When an agent receives a task request from a user, it first utilizes the LLM's planning capability to generate a plan, decomposing a complex task into a series of simple sub-tasks. Then, by invoking tools such as APIs, code interpreters, and search engines, and combining them with the memory possessed by the LLM (such as user personal preferences), it executes the sub-tasks to complete the entire mission. 

For example, when a user submits the request "Book a ticket from Hefei to Beijing for the day after tomorrow," the plan generated by the agent using the LLM might be: 
\begin{enumerate}
    \item Query the date for the day after tomorrow;
    \item Query the weather;
    \item Query flight schedules and prices from Hefei to Beijing;
    \item Query train schedules and prices from Hefei to Beijing;
    \item Compare flight and train prices;
    \item Query the user's personal schedule;
    \item Combine the above information and user preferences to recommend a travel plan.
\end{enumerate}
After deriving the plan, the agent completes sub-tasks 1, 2, 3, 4, and 6 by calling applications on the mobile phone or open APIs; completes sub-task 5 via a code interpreter; and finally, for sub-task 7, combines weather, price information, personal schedule, and personal preferences stored in memory to obtain the travel plan.

In this example, we can see that agents integrated with LLM capabilities can better complete tasks by combining user personal needs \cite{xi2023rise, openassistant2024}. However, the computation and latency costs of using LLMs cannot be ignored. Using AutoGen \cite{wu2023autogen} as the agent framework and OpenAI's GPT-4 API as the LLM, for the input task "Book a ticket from Hefei to Beijing for the day after tomorrow," the latency for returning the plan is approximately 25 seconds, consuming 2015 tokens, with a cost of \$0.073 (approx. 0.53 RMB). Within the 25-second latency, the plan generation latency is about 24 seconds, with the rest being network latency. This is because LLMs generate plans in an auto-regressive manner, and plan texts are generally long. A plan generation latency of 25 seconds, plus the subsequent execution latency, significantly affects user experience. Analysis of the personal assistant dataset in  \cite{gill2024privacy} and  \cite{openassistant2024} indicates that about 30\% of task requests in LLM-driven agents are semantically identical or similar. Therefore, caching the plans generated by the agent using the LLM and reusing previous plans when identical or similar requests arrive subsequently is an effective method to reduce the response latency of LLM-driven agents \cite{zhao2023survey}.

Zhou et al. \cite{fu2023gptcache} studied caching and reuse methods for the responses of the LLM itself. By vectorizing request texts (embedding) and comparing the semantic similarity between different request texts, it judges whether a subsequent request can reuse the response of a previous request. For example, "Please explain what AIoT is" and "What is AIoT" are semantically similar; one can directly reuse the LLM's response to the former without generating a new introduction, thereby reducing latency. 

However, in the process of an LLM-driven agent handling user requests, consider the following two requests: 
\begin{itemize}
    \item Request 1: "Book a ticket from Hefei to Beijing for the day after tomorrow"
    \item Request 2: "Book a ticket from Changsha to Shanghai for tomorrow"
\end{itemize}
If GPTCache\cite{fu2023gptcache} is directly adopted, and Request 2 is judged similar to Request 1, directly returning Request 1's response would obviously provide wrong ticket information to the user. If Request 2 is judged dissimilar to Request 1, it results in a plan generation latency of about 25 seconds. In fact, upon observation, it is not difficult to find that for Request 1 and Request 2, although the time, origin, and destination differ, the \textit{task type} is the same. Therefore, the plan steps produced by the agent should be the same, with only the specific key parameters ("Time, Origin, Destination") being different. That is, Request 2 should be judged as similar to Request 1, but instead of returning Request 1's response, Request 2 should reuse the \textit{plan} generated by the LLM for Request 1.

This paper conducts research on a plan reuse mechanism for LLM-driven agents. Existing research mainly focuses on the caching and reuse mechanisms for the LLM's direct responses. Our research faces three main challenges:

\textbf{Challenge 1: How to define if requests are similar.} As mentioned, in response caching methods, semantic similarity is defined by vectorizing original texts. However, in LLM-driven agents, Request 1 and Request 2 have different key parameters. Tests show that calculating similarity on raw text (with a threshold of 0.75) judges them as dissimilar, whereas Request 2 can factually reuse the plan for Request 1. Thus, a new method is needed to define similarity between request texts in LLM-driven agents.

\textbf{Challenge 2: How to identify key parameters.} After defining similarity, if two requests are similar, the later request can reuse the plan of the former. During reuse, key parameters must be replaced. In Request 1 and 2, key parameters are time (day after tomorrow vs. tomorrow), origin (Hefei vs. Changsha), and destination (Beijing vs. Shanghai). Intuitively, string matching could replace parameters, but natural language contains synonyms and varying word orders (e.g., Request 2 could be "Book me a ticket, Changsha to Shanghai, tomorrow"), which string matching cannot easily handle. Therefore, a method capable of semantic-aware parameter identification is required.

\textbf{Challenge 3: How to correctly and effectively reuse plans.} Currently, plans output by agents are usually unstructured text. Even if key parameters are identified, it is difficult to determine which sub-task each parameter should apply to, making effective plan reuse hard. Therefore, a method to convert unstructured text into structured execution plans is needed.

To address the above challenges, this paper proposes \textbf{AgentReuse}, a plan reuse mechanism for LLM-driven agents, and implements it with millisecond-level extra latency. The main contributions of this paper are in three aspects:

\begin{enumerate}
    \item Addressing the long response time of LLM-driven agents, we formally describe and analyze the workflow of LLM-driven agents and propose a research approach for reusing plans generated by such agents.
    \item We propose and implement AgentReuse with low overhead. AgentReuse identifies key parameters by leveraging the semantic differences between request texts. After extracting key parameters, it uses an intent classification-based method to define similarity between user requests using semantic similarity, enabling plan reuse.
    \item To verify the effectiveness of AgentReuse, we conducted experiments using a real-world dataset. The results show that compared to existing methods, AgentReuse achieves an effective plan reuse rate of 93\%. The F1 score for similarity definition increased by 6.8 percentage points, accuracy increased by 13.06 percentage points, and latency was reduced by 93.12\%.
\end{enumerate}
\section{Background and Related Work}\label{sec:back}
Focusing on the research theme of this paper, this section briefly introduces the background knowledge and recent research progress in two aspects: LLM-driven agents and caching \& reuse methods.

\subsection{LLM-driven Agents}
An agent is an autonomous algorithmic system capable of environmental perception, decision-making, and action execution, and has always been a key focus of research in both academia and industry. The original intention of developing agents was to simulate the intelligent behavior of humans or other organisms, aiming to solve problems or execute tasks automatically. However, traditional agents face major challenges as they usually rely on heuristic rules or are limited to specific environmental constraints, which largely restricts their adaptability and scalability in open and dynamic scenarios \cite{lin2024parrot}.

Since Large Language Models (LLMs) have demonstrated excellent capabilities in solving complex tasks, an increasing number of research works have begun to explore using LLMs as the control center for agents to improve their performance in open domains and dynamic environments \cite{mindstream2024autogpt}, such as the series of LLM-driven agents represented by AutoGPT \cite{mindstream2024autogpt}, AutoGen \cite{wu2023autogen}, HuggingGPT \cite{shen2023hugginggpt}, and MetaGPT \cite{hong2023metagpt}.

For LLM-driven agents, researchers propose that \textit{Agent = LLM + Memory + Planning + Tool Use}. That is, using the LLM as the core controller, supplemented by capabilities such as planning, memory, and tool usage, the agent completes tasks submitted by users by coordinating these different capabilities \cite{wang2024survey}. Hong et al. \cite{hong2023metagpt} designed MetaGPT, which assigns divisions of labor to multiple agents to jointly complete the development and debugging of computer software. Bram et al. \cite{bram2024augmenting} proposed ChemCrow, which integrates 17 tools designed by experts to enhance the capabilities of LLMs in the field of chemistry.  ResearchAgent \cite{baek2024researchagent} mimics the steps of humans writing scientific research papers, guiding the LLM to generate complete research ideas including problem identification, method development, and experimental design.

This paper conducts research focusing on typical LLM-driven AIoT applications, namely personal agents. Regarding personal agents, VIVO released Blue Heart V based on the BlueLM \cite{vivo2024bluelm}; Xiaomi's Xiao Ai has also integrated LLMs \cite{xiaomi_xiaoai}; Google's Pixel 8 Pro \cite{google2024pixel} and Bard assistant \cite{hsiao2024assistant} have both integrated LLMs; and Microsoft has integrated LLM capabilities into Microsoft Copilot \cite{microsoft2024copilot} and launched Copilot+PC \cite{mehdi2024copilotpc}. Furthermore,  AppAgent \cite{zhang2023appagent} and AutoDroid \cite{wen2024autodroid} can execute various tasks on mobile phones by autonomously learning and mimicking human clicking and sliding gestures. This paper mainly focuses on personal agents, aiming to reduce agent response latency by analyzing the similarity between task request texts submitted by users and reusing plans generated by the LLM.

\subsection{Caching and Reuse Methods}
In computer architecture, caching aims to bridge the gap between memory access latency and processor processing speed \cite{sedaghati2022xcache, bhatla2024maya}. Today, the concept of caching has expanded to various fields and plays an important role in improving computer system performance. Structures located between different types of devices to eliminate the impact of access time differences can be referred to as caches, such as disk caches \cite{wong2024baleen, liu2024optimizing, mcallister2023kangaroo}, CDN caches \cite{chen2024darwin, yang2023fifo, yan2022towards}, Web caches \cite{mirheidari2020cached, wang2020rldish}, and container caches \cite{fuerst2021faascache, roy2022icebreaker, li2024online}, etc. The effectiveness of caching stems from the temporal locality of requests, meaning that in computer applications, content that has just been requested is likely to be requested again \cite{traverso2013temporal}. Caching recently requested content for reuse in the next request can effectively reduce response latency.

In the past two years, researchers have explored the application of caching and reuse technologies in LLMs. Researchers have conducted research on KV Cache in LLMs. 
Based on the calculation characteristics of Self-Attention, Kwon et al. \cite{kwon2023efficient} proposed PagedAttention, a KV Cache management method. Zhang et al. \cite{zhang2023h2o} introduced H2O to compress KV Cache based on the varying importance of tokens, while Liu et al. \cite{liu2023cachegen} proposed CacheGen to perform dynamic recomputation of KV Cache, addressing system-level challenges when LLMs process long contexts. HotPrefix~\cite{li2025hotprefix}  and IMPRESS~\cite{chen2025impress}  further advance KV cache optimization, particularly for scenarios involving shared long contexts or system prompts.   Furthermore, several studies \cite{gill2024privacy, zhu2023optimal, fu2023gptcache} have focused on caching and reusing LLM-generated responses based on semantic similarity between request texts. Specifically, Gill et al. \cite{gill2024privacy} proposed a caching and reuse method based on federated learning that considers user privacy and similarity retrieval costs. Zhu et al. \cite{zhu2023optimal} provided a theoretical analysis of caching mechanisms for LLM responses, and Fu and Feng \cite{fu2023gptcache} developed open-source tools enabling developers to integrate caching and reuse mechanisms into LLMs. In the domain of text-to-image diffusion models, the NIRVANA method \cite{agarwal2024approximate} reduces iterative denoising steps by reusing intermediate noise states created during previous image generation

This paper mainly focuses on caching and reusing the \textit{plans} generated by LLM-driven agents. Existing methods mainly propose caching and reuse methods for the \textit{responses} of LLMs, which cannot be directly applied to LLM-driven agents.

\section{Workflow and Research Objectives of LLM-Driven Agents}
This section formally describes the workflow of the LLM-driven agent studied in this paper and briefly introduces the research objectives.

\begin{figure}[h]
    \centering
    \includegraphics[width=0.95\linewidth]{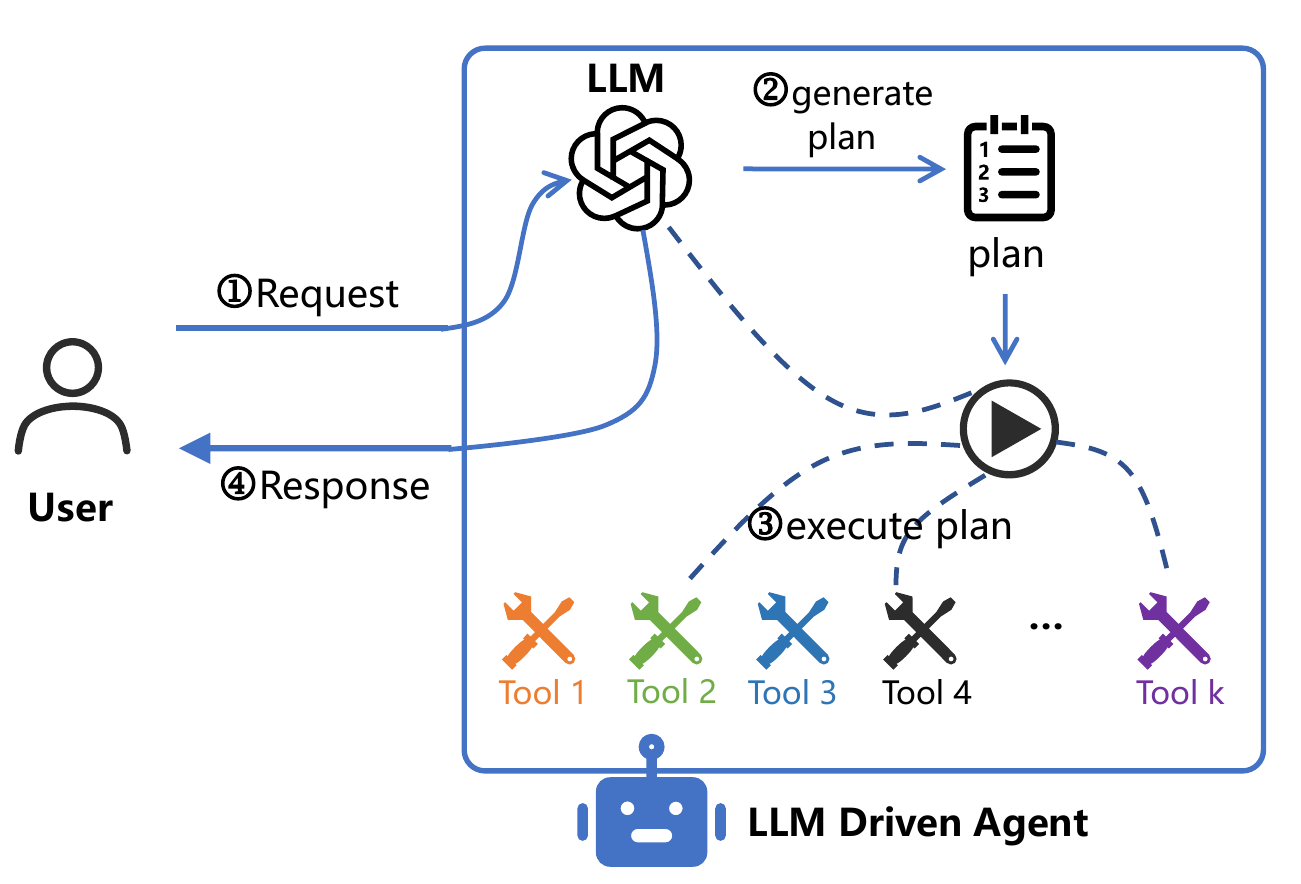}
    \caption{Workflow of LLM-driven agent}
    \label{fig:workflow}
\end{figure}

As shown in Figure \ref{fig:workflow}, when the LLM-driven agent studied in this paper serves a user:
\begin{enumerate}
    \item The agent receives a task request $q_i$ submitted by the user, which is a natural language text.
    \item Upon receiving the request, the agent uses the planning capability of the LLM to generate a plan $p_i$ for $q_i$.
    \item After obtaining the plan $p_i$, the agent executes the plan using various tools such as search engines and code interpreters to obtain the response $r_i$. The response $r_i$ can be the natural language text desired by the user or a response code returned after task execution (e.g., a 200 code in HTTP services).
    \item The agent returns the response to the user.
\end{enumerate}

Without loss of generality, we define $Q = \{q_1, q_2, \dots\}$, $P = \{p_1, p_2, \dots\}$, and $R = \{r_1, r_2, \dots\}$. For request $q_i$, the time taken to generate the plan text is referred to as the \textit{plan generation latency}, denoted as $t_p$. Specifically, $t_p$ refers to the time interval from the moment the agent receives $q_i$ to the moment the plan text $p_i$ generation is completed (i.e., generating the \texttt{<EOS>} token). The time taken to execute the plan is referred to as the \textit{execution latency}, denoted as $t_e$. This refers to the time interval from the moment the plan text $p_i$ generation is completed to the moment the response $r_i$ is produced. The total response latency for request $q_i$ is $t = t_p + t_e$.

As mentioned earlier, among the task requests received by the agent, approximately 30\% are semantically identical or similar. If we can identify identical or similar requests and reuse the plans generated by the agent for previous requests, we can eliminate the plan generation latency $t_p$, thereby reducing the response latency and optimizing the user experience. In this paper, requests that can factually reuse previous plans are defined as \textit{reusable requests}, and requests that factually cannot reuse previous plans are defined as \textit{non-reusable requests}.

When defining the similarity between requests, if a non-reusable request is judged as reusable, it will lead to the reuse of an incorrect plan, consequently returning a wrong response to the user. Conversely, if a reusable request is judged as non-reusable, it incurs extra plan generation latency. Therefore, designing a method that can accurately define the similarity between requests is essential.

Furthermore, for reusable requests, when reusing a previously generated plan, certain parameters in the original plan need to be replaced with the relevant parameters from the current request. Currently, there is no method that effectively achieves parameter replacement; that is, plan reuse cannot be effectively completed. Thus, we also need to design a method capable of effectively reusing plans.

This section mainly introduced the workflow of LLM-driven agents, including four parts: receiving user requests, generating plans, executing plans, and returning responses, and provided a formal description. Additionally, this section defined reusable and non-reusable requests, clarifying the importance of designing methods for accurate similarity definition and effective plan reuse.

\section{Reuse Mechanism Design}\label{sec:des}

This section introduces the specific content of the reuse mechanism for LLM-driven agents proposed in this paper, \textbf{AgentReuse}, and explains the workflow of AgentReuse using examples.

When an LLM-driven agent serves a user, it receives a request $q_i$, utilizes the LLM's planning capability to obtain a plan text $p_i$, and finally uses tools to execute the plan to get a response $r_i$. The plan is generated by the LLM, and the plan generation latency $t_p$ is proportional to the length of the plan text; the longer the plan text, the larger $t_p$. Using AutoGen \cite{wu2023autogen} as the agent framework and OpenAI's GPT-4 API as the LLM, 100 tests were conducted on a real-world dataset \cite{openassistant2024}. The results indicate that the average plan generation latency is 31.8s. Previous research \cite{gill2024privacy} and our tests on the real-world dataset both indicate that about 30\% of requests are identical or similar. For identical or similar requests, if the plan or response of a previous request can be reused, the response latency can be reduced to a certain extent. Based on this, our goal is to design a plan reuse mechanism for LLM-driven agents to reduce response latency.

\begin{figure}[h]
    \centering
    \includegraphics[width=0.95\linewidth]{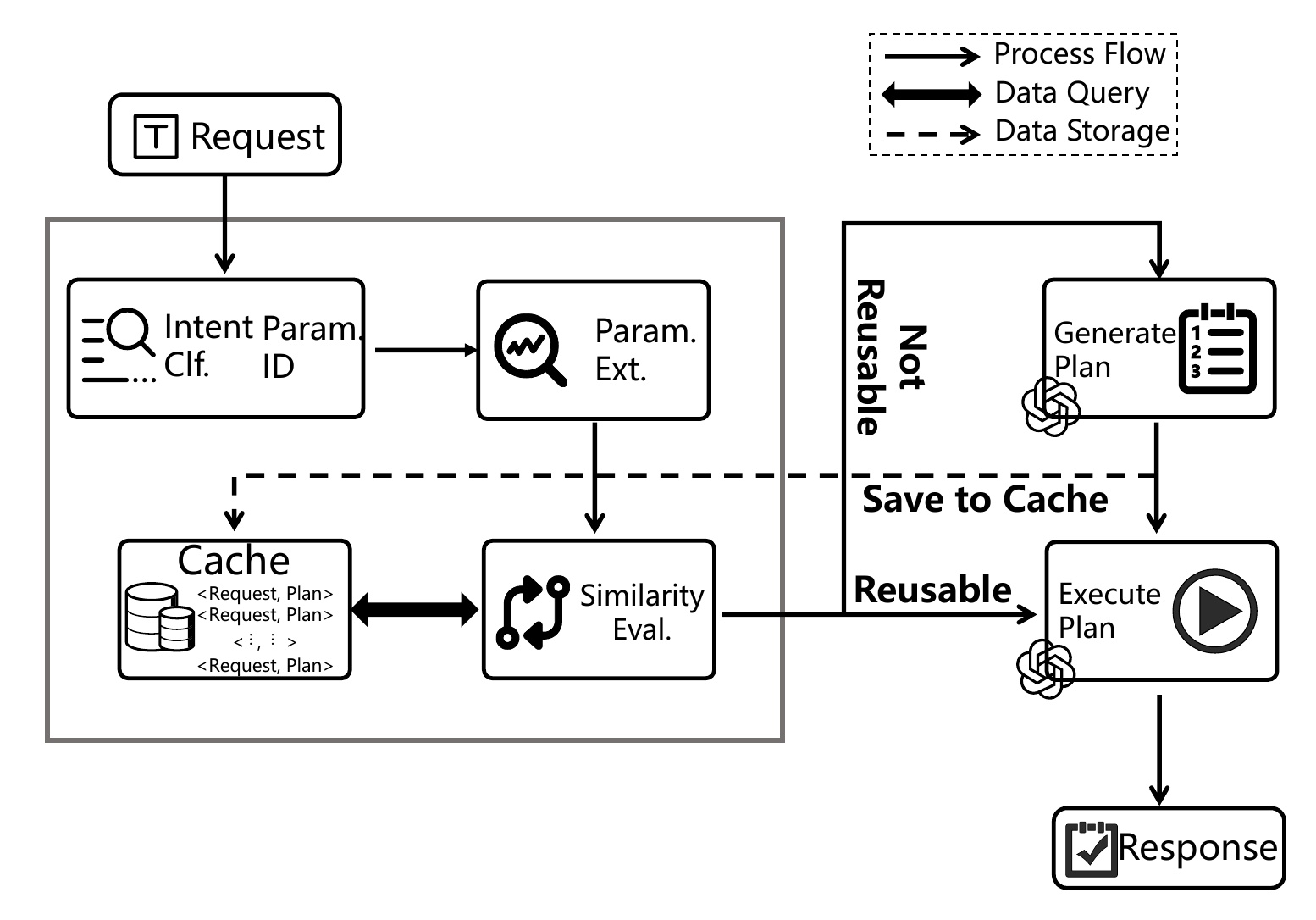}
    \caption{Flowchart of AgentReuse}
    \label{fig:flowchart}
\end{figure}

As shown in Algorithm \ref{alg:agentreuse} and Figure \ref{fig:flowchart}, AgentReuse utilizes the semantic similarity and differences between request texts to effectively reuse the plans of LLM-driven agents.

\begin{algorithm}[h]
\caption{AgentReuse}
\label{alg:agentreuse}
\begin{algorithmic}[1]
\REQUIRE Requests $Q = \{q_1, q_2, \dots\}$, Cache $\{c_0, c_1, \dots, c_n\}$, Intent Categories, Similarity threshold $\gamma$.
\ENSURE Response $R = \{r_1, r_2, \dots\}$.
\STATE Initialize Cache as empty.
\FOR{each $q_i$}
    \STATE Vectorize $q_i$ to get $E(q_i)$.
    \STATE Perform intent classification and parameter identification on $q_i$, obtaining intent $c_i$ and parameters $para_i$.
    \IF{$c_i \neq$ Undefined Intent}
        \STATE Extract parameters $para_i$ from $q_i$ to get $q_{i}^{-}$ and $E(q_{i}^{-})$.
        \STATE Perform similarity search for $E(q_{i}^{-})$ in Cache (within category $c_i$).
        \IF{Highest similarity score $\geq \gamma$}
            \STATE Reuse plan $p_m$ from $q_m$ (where $E(q_{m}^{-})$ has the highest similarity with $E(q_{i}^{-})$ and $\ge \gamma$).
            \STATE Execute plan $p_m$ using tools with parameters $para_i$ to get response $r_i$.
        \ELSE
            \STATE Call LLM to generate plan $p_i$.
            \STATE Execute plan using tools to get response $r_i$.
            \STATE Store $\langle E(q_{i}^{-}), p_i, c_i \rangle$ into Cache.
        \ENDIF
    \ELSE
        \STATE Call LLM to generate plan $p_i$.
        \STATE Execute plan using tools to get response $r_i$.
    \ENDIF
\ENDFOR
\RETURN Response $R$.
\end{algorithmic}
\end{algorithm}

Next, we explain AgentReuse using the example $q_i =$ "Book a ticket from Hefei to Beijing for the day after tomorrow." First, we vectorize the request text $q_i$ and perform intent classification. Intent classification refers to classifying request texts into different categories based on their semantics, utilizing semantic similarities between different texts to group them into the same intent category. In personal assistants, intent categories mainly include coarse classifications such as query, play, send, download, and book. The intent of the current example $q_i$ is "Book." Furthermore, slot filling tasks are usually performed jointly with intent classification tasks \cite{ma2024twostage}. Slot filling is used to identify key parameters and temporarily store the parameters "(day after tomorrow, Hefei, Beijing)" into $para_i$. Subsequently, within the intent category "Book," parameter extraction and similarity definition are performed. For parameter extraction, in the example, the parameters "(day after tomorrow, Hefei, Beijing)" need to be extracted. Removing these parameters from the request text $q_i$ yields $q_{i}^{-} =$ "Book a ticket from to". This allows it to be distinguished from tasks like booking restaurants or meeting rooms within the "Book" class, finding requests more similar to "Book a ticket from to" and reusing the plans of similar requests. Similarity is defined by the similarity threshold $\gamma$. $\gamma$ is an empirical value, and this paper sets the default value of $\gamma$ to 0.75.

To effectively reuse plans, when reusing a plan, the previously identified parameters $para_i$ are used as input, utilizing the semantic differences between request texts. When there is no request similar to $q_{i}^{-}$ in the cache, the LLM is called to generate a plan, which is then executed using tools to get response $r_i$, and $\langle E(q_{i}^{-}), p_i, c_i \rangle$ is stored in the cache. Here, $E(q_{i}^{-})$ is the vectorized $q_{i}^{-}$, $p_i$ is the plan generated by the LLM, and $c_i$ is the intent category of text $q_i$. When $q_i$ is not within the "Book" intent category, i.e., the intent category is undefined, the LLM is directly called to generate a plan $p_i$, which is executed to get the response $r_i$. It is worth noting that in Figure \ref{fig:flowchart}, parameter extraction, similarity definition, and cache lookup are performed within the current intent category, which can reduce search costs.

In AgentReuse, this paper classifies the intent of user requests by utilizing the semantic similarity between request texts, while simultaneously utilizing the semantic differences between request texts to identify and extract key parameters. We use the identified key parameters to perform similarity evaluation and parameter replacement on request texts, achieving accurate similarity definition for requests received by LLM-driven agents and effective reuse of plans. Additionally, since requests submitted by users may involve real-time changing information, such as flight information, this paper reuses the \textit{plan} rather than the \textit{response}.

Compared to the execution process of existing LLM-driven agents, AgentReuse incurs additional costs in terms of storage and latency.
\begin{enumerate}
    \item \textbf{Storage Cost:} AgentReuse needs to use models to vectorize texts and classify intents. When the model is running, it occupies VRAM. Vectorized texts and plans generated by the LLM need to be cached, occupying storage space.
    \item \textbf{Latency Cost:} Compared to the existing execution process, the new workflow introduced by AgentReuse brings extra runtime latency. The significant parts of the latency are using models for intent classification and parameter identification, as well as defining the similarity between user requests and cached requests.
\end{enumerate}

It is worth noting that existing personal assistants already need to use models to classify user request intents \cite{nvidia_iva, techtarget_va}. Therefore, the extra storage and latency costs brought by intent classification and parameter identification may not be counted towards the cost of AgentReuse. A specific analysis and discussion of the extra costs will be conducted in Section \ref{sec:exp}.

This section introduced the core content of this paper, the plan reuse mechanism AgentReuse for LLM-driven agents. To define the similarity of requests, AgentReuse performs intent classification and key parameter identification on requests, extracts key parameters, and then performs similarity definition to improve performance. If a reusable plan exists in the cache, it is reused and executed; if not, a new plan is generated and stored in the cache. Finally, this section briefly analyzed the additional costs incurred by AgentReuse. To effectively reuse plans, this paper adopts a method of adding prompt words to obtain structured plans, which will be introduced in the next section.

\section{Implementation of the Reuse Mechanism} \label{sec:imple}
Based on the plan reuse mechanism for LLM-driven agents, \textbf{AgentReuse}, proposed in the previous section, this section aims to introduce the relevant technologies for implementing AgentReuse. It is worth noting that the relevant technologies adopted in this section represent only one method for implementing AgentReuse. For different performance requirements, extra cost limitations, and input text language types, different technical routes can be selected.

\subsection{Models and Techniques}
For \IntentClf and \ParamID, this paper adopts the \texttt{bert-base-chinese} model \cite{googlebert}. This model is the Chinese version of the BERT model and possesses powerful language understanding capabilities. The model uses a bidirectional Transformer architecture and is pre-trained on a large amount of Chinese corpus data. It performs excellently on multiple Chinese natural language processing tasks, such as text classification, named entity recognition, and sentiment analysis, providing a solid foundation for Chinese NLP tasks.

For \textbf{Text Vectorization}, this paper adopts the \texttt{m3e-small} model \cite{mokaai}. M3E refers to Moka Massive Mixed Embedding. Specifically, this model is trained by MokaAI using a massive dataset of Chinese sentence pairs and supports mixed Chinese-English homogeneous text similarity calculation. It is a text embedding model capable of converting natural language into dense vectors. The \texttt{m3e-small} model has 24 million parameters and generates vectors with a dimension of 512.

For \textbf{Text Semantic Similarity Calculation}, this paper employs Cosine Similarity. Vectors can capture the semantic meaning of words or sentences; words or sentences with similar meanings will have similar vector representations. For example, "train" and "high-speed rail," although textually different, are semantically similar. Therefore, after vectorization, the vectors corresponding to these two terms are close in the vector space. Cosine similarity is typically used to quantify the similarity between vectors, calculated by taking the cosine of the angle between two vectors. The range of cosine similarity is -1 to 1, where -1 indicates the two vectors are completely opposite in direction, 0 indicates they are orthogonal, and 1 indicates they are in the same direction. Generally, in text processing, the closer the cosine similarity value is to 1, the more similar the two texts are. For vectors $\mathbf{V}_1$ and $\mathbf{V}_2$, the cosine similarity is:

\begin{equation}
    \cos \theta = \frac{\mathbf{V}_1 \cdot \mathbf{V}_2}{||\mathbf{V}_1|| \cdot ||\mathbf{V}_2||}
    \label{eq:cosine}
\end{equation}

For \textbf{Vector Similarity Search}, in the technical implementation of this paper, we use the \texttt{IndexFlatIP} index from Facebook AI Similarity Search (FAISS) \cite{faiss} for efficient similarity search. FAISS is an open-source similarity search library by Facebook that provides efficient similarity search and clustering for vectors, supporting searches at the billion-scale level, and is currently the most mature approximate nearest neighbor search library. It is worth noting that FAISS does not directly provide cosine similarity calculation; the \texttt{IndexFlatIP} selected in this paper performs dot product calculation. Therefore, when performing cosine similarity calculation, the vectors need to be normalized first.

\subsection{Structured Plan Reuse}
Furthermore, since plans generated by LLM-driven agents like AutoGen are usually natural language texts, to store and effectively reuse plans, this paper performs structured processing on the natural language text plans. In the plan generation prompt words used by AutoGen with the LLM, we prompt the LLM to simply annotate the dependency information between sub-tasks when generating the plan. Based on the dependency annotations, the plan text can be processed to obtain a structured plan. Additionally, in agent frameworks like AutoGen, for system security considerations, tool usage is usually conducted within containers \cite{patil2024goex, chu2024tutorial, langchain2024security}. Therefore, container image information needs to be recorded in the steps for subsequent reuse. 

For the natural language text plan $p_i$ of request $q_i$, the plan is represented as:
\begin{itemize}
    \item Step 1, Step Description, Container Image, Input, Dependency, Output;
    \item Step 2, Step Description, Container Image, Input, Dependency, Output;
    \item ...
    \item Step $k$, Step Description, Container Image, Input, Dependency, Output.
\end{itemize}

Taking the plan generated by "Book a ticket from Hefei to Beijing for the day after tomorrow" as an example, this paper converts it into the following structure (using the abbreviations requested):

\begin{description}
    \item[Step 1] \textbf{\DateQuery}, \texttt{query\_date}, Date Desc., No Dependency, Output: Date.
    \item[Step 2] \textbf{\WeatherQuery}, \texttt{query\_weather}, (Location, Date), Dep: Step 1, Output: Weather.
    \item[Step 3] \textbf{\FlightQuery}, \texttt{query\_flight}, (Origin, Dest, Date), Dep: Step 1, Output: Flight Info.
    \item[Step 4] \textbf{\TrainQuery}, \texttt{query\_train}, (Origin, Dest, Date), Dep: Step 1, Output: Train Info.
    \item[Step 5] \textbf{\PriceComp}, \texttt{compare\_price}, (Flight Info, Train Info), Dep: Step 3, 4, Output: Comparison.
    \item[Step 6] \textbf{\ScheduleQuery}, \texttt{query\_schedule}, Date, Dep: Step 1, Output: Personal Schedule.
    \item[Step 7] \textbf{\PrefQuery} (Recommend Travel Plan), \texttt{call\_prefer}, All Info, Dep: Step 2, 5, 6, Output: Recommendation.
\end{description}

\begin{figure}[h]
    \centering
    \includegraphics[width=0.8\linewidth]{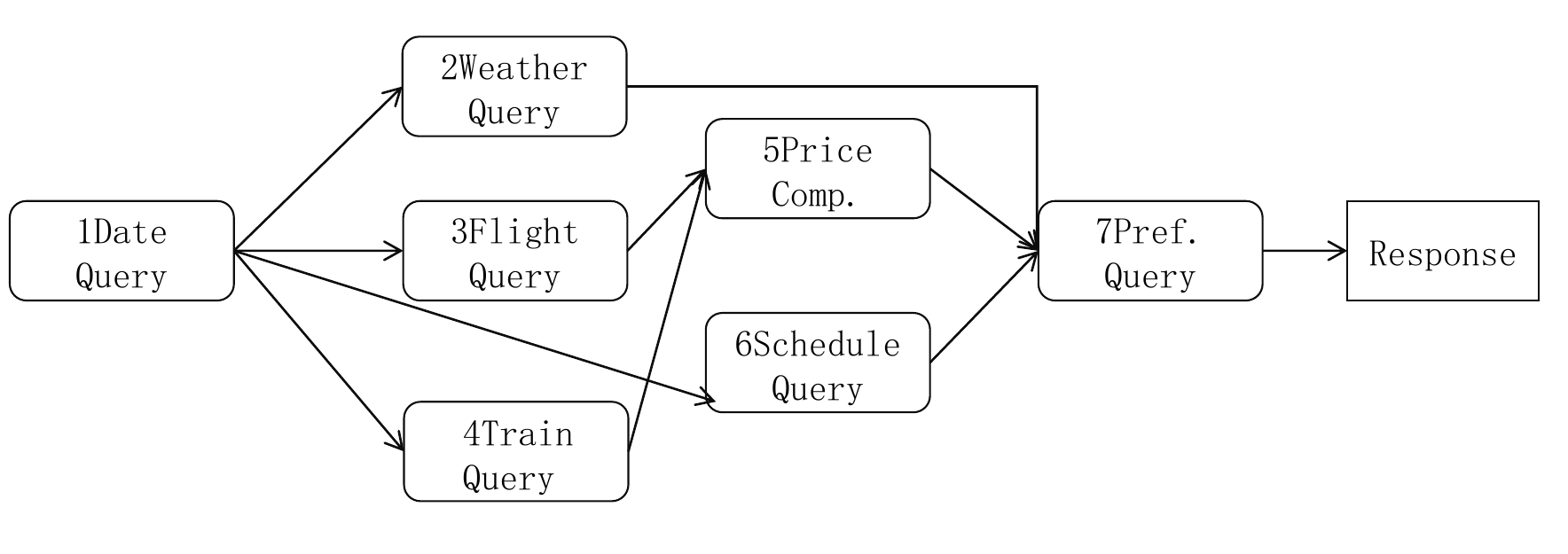}
    \caption{Execution graph for the plan generated by the request}
    \label{fig:exec_graph}
\end{figure}

Figure \ref{fig:exec_graph} illustrates the execution graph of the above plan. For each step, the current step can only be executed after its dependent steps are completed. Specifically, Steps 2, 3, 4, and 6 depend on Step 1; Step 5 depends on Steps 3 and 4; and Step 7 depends on Steps 2, 5, and 6. In actual execution, the steps are implemented as follows: use open APIs to complete Steps 1 and 2; use open APIs or applications like Ctrip and 12306 to complete Steps 3 and 4; use Python code to complete Step 5; use a calendar application to complete Step 6; and finally, invoke the LLM to complete Step 7.

This section introduced the technical methods adopted to implement AgentReuse, including using \texttt{bert-base-chinese} as the intent recognition model, \texttt{m3e-small} as the text vectorization model, and adopting cosine similarity to evaluate text semantic similarity. As mentioned at the beginning of this section, the methods introduced here are just one of the schemes to implement AgentReuse. For different languages, different performance requirements, and different resource usage limits, different models and related technologies can be selected.

\section{Experiments and Result Analysis}\label{sec:exp}
To evaluate the effectiveness and extra costs of the proposed mechanism AgentReuse, we conducted a series of experiments based on a real-world dataset. This section first details the experimental setup, then introduces the evaluation metrics, and finally presents the experimental results and analyzes the extra costs. Experimental results show that compared to existing methods, AgentReuse achieves effective plan reuse with an effective reuse rate of 93\%. Its F1 score for similarity definition is 6.8 percentage points higher than existing methods, accuracy is 13.06 percentage points higher, and latency is reduced by 60.61 percentage points. Compared to the execution process of existing LLM-driven personal agents, AgentReuse incurs an extra VRAM usage of about 100 MB, memory usage for processing each request does not exceed 1 MB, and latency does not exceed 10 ms.

\subsection{Experimental Setup}
The experiments in this paper use the SMP dataset \cite{smp2019}, which contains 2,664 task requests for LLM-driven personal agents. Examples include text requests like "What are the flight times to Chengdu tomorrow morning" and "Play a poem by Li Qingzhao," covering 23 intent categories such as "LAUNCH," "QUERY," and "ROUTE."

The experiments were conducted on a server with the following configuration: Operating System Ubuntu 20.04.6 LTS, Intel Xeon Gold 6330 CPU @ 2.00 GHz CPU , GPU NVIDIA A100 80 GB, Python version 3.10, CUDA version 11.4, and server memory of 503 GB.

To test the effectiveness of AgentReuse, this paper selected two methods proposed by other researchers as comparison methods (Baselines):
\begin{enumerate}
    \item \textbf{GPTCache} \cite{fu2023gptcache}: A mechanism for caching and reusing LLM responses. When defining similarity, it directly vectorizes the original request text to calculate semantic similarity.
    \item \textbf{MeanCache} \cite{gill2024privacy}: A mechanism for caching and reusing LLM responses. It performs Principal Component Analysis (PCA) on the request text to reduce dimensions to 64 before performing similarity definition.
\end{enumerate}

In addition, to verify the effectiveness of different modules in AgentReuse (i.e., ablation studies), two variants were tested:
\begin{enumerate}
    \item \textbf{OneIntent}: When defining similarity in the implemented AgentReuse, retrieval is performed on all cached content instead of only within the current intent category.
    \item \textbf{WithArgs}: When defining similarity in the implemented AgentReuse, parameter extraction is not performed; instead, the vectorized original request text is directly used for similarity search.
\end{enumerate}

\subsection{Evaluation Metrics}
We need to evaluate the performance of AgentReuse's similarity definition, the effectiveness of plan reuse, extra costs, and performance gains. This section introduces the metrics used to evaluate the above experimental effects respectively.

To evaluate the performance of semantic similarity identification in our caching system, we employ four standard metrics: Precision, Recall, $F_{\beta}$ Score (specifically the F1 Score), and Accuracy. These metrics are calculated based on the classification of request reusability into True Positives (TP), False Positives (FP), True Negatives (TN), and False Negatives (FN). While Precision and Recall measure the system's ability to correctly identify reusable requests and minimize false identifications, the F1 Score and Accuracy provide a comprehensive assessment of the model's overall predictive reliability.

This paper uses the effective reuse rate as a quantitative metric to evaluate the effectiveness of plan reuse, detailed in Section ~\ref{sec:subres}.

For the extra costs of using AgentReuse, this paper uses the extra latency per request and the extra memory occupied (including VRAM) as evaluation metrics.

For the assessment of performance gains, since the use of different reuse mechanisms has no impact on the latency of \textit{executing} the plan, the analysis of performance gains in this section only considers the plan \textit{generation} latency and the extra costs of using the mechanism. It is assumed that all non-True Positive cases require a plan generation.

\subsection{Result Analysis}\label{sec:subres}
This section mainly introduces the experimental results of similarity definition and plan reuse effectiveness, and analyzes the results.

First, we analyze the experimental results of similarity definition. Only when the similarity definition result is accurate can plan reuse be effective; otherwise, it may lead to consequences such as reusable requests not being judged as reusable, or non-reusable requests being judged as reusable (reusing previous plans), leading to incorrect responses.

Table \ref{tab:results} shows the similarity definition experimental results under the default setting, i.e., similarity threshold $\gamma = 0.75$. The proposed mechanism AgentReuse achieves the highest F1 Score and Accuracy, which are 0.9718 and 0.9459, respectively. Among the comparison methods, the OneIntent variant (which searches all cached content during similarity retrieval instead of just within the current intent category) has performance very close to AgentReuse, but its runtime latency is 17.66\% slower than AgentReuse (extra latency costs are discussed later). Compared to GPTCache, which targets LLM response reuse, AgentReuse improves the F1 Score and Accuracy by 6.8 and 13.06 percentage points, respectively.

\begin{table}[h]
\centering
\caption{Experimental Results for Similarity Evaluation ($\gamma = 0.75$)}
\label{tab:results}
\begin{tabular}{lcccc}
\hline
Method & F1 Score & Precision & Recall & Accuracy \\
\hline
GPTCache & 0.9100 & 0.9887 & 0.8428 & 0.8366 \\
MeanCache & 0.6211 & 0.9991 & 0.4506 & 0.4614 \\
OneIntent & 0.9712 & 0.9857 & 0.9571 & 0.9444 \\
WithArgs & 0.9035 & 0.9967 & 0.8262 & 0.8272 \\
\textbf{AgentReuse} & \textbf{0.9718} & 0.9931 & 0.9513 & \textbf{0.9459} \\
\hline
\end{tabular}
\end{table}

\begin{figure}[h]
\centering
    \includegraphics[width=0.8\linewidth]{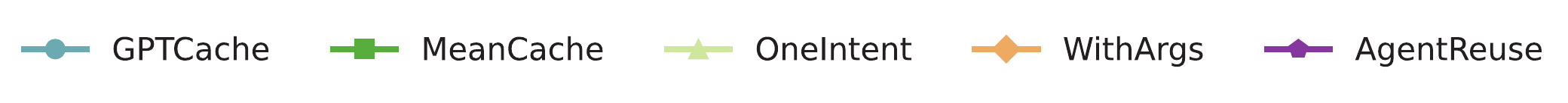}
    \centering
    
    \begin{subfigure}[t]{0.48\linewidth}
        \centering
        \includegraphics[width=\linewidth]{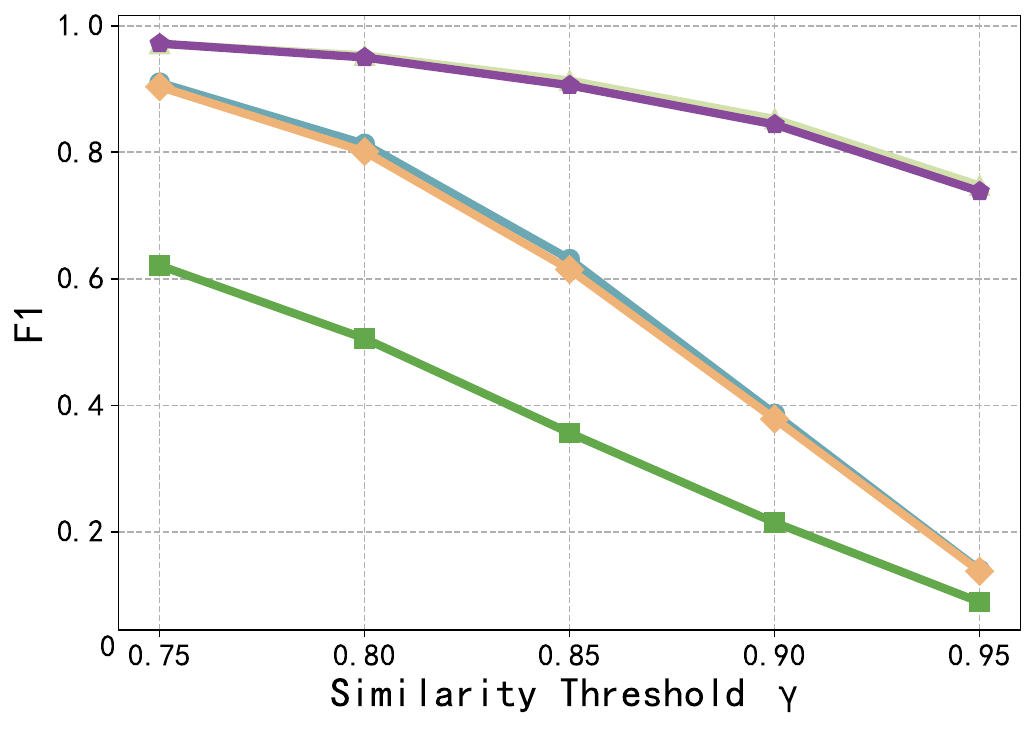}
        \caption{F1 vs $\gamma$}
        \label{fig:f1_vs_gamma}
    \end{subfigure}
    \hfill
    \begin{subfigure}[t]{0.48\linewidth}
        \centering
        \includegraphics[width=\linewidth]{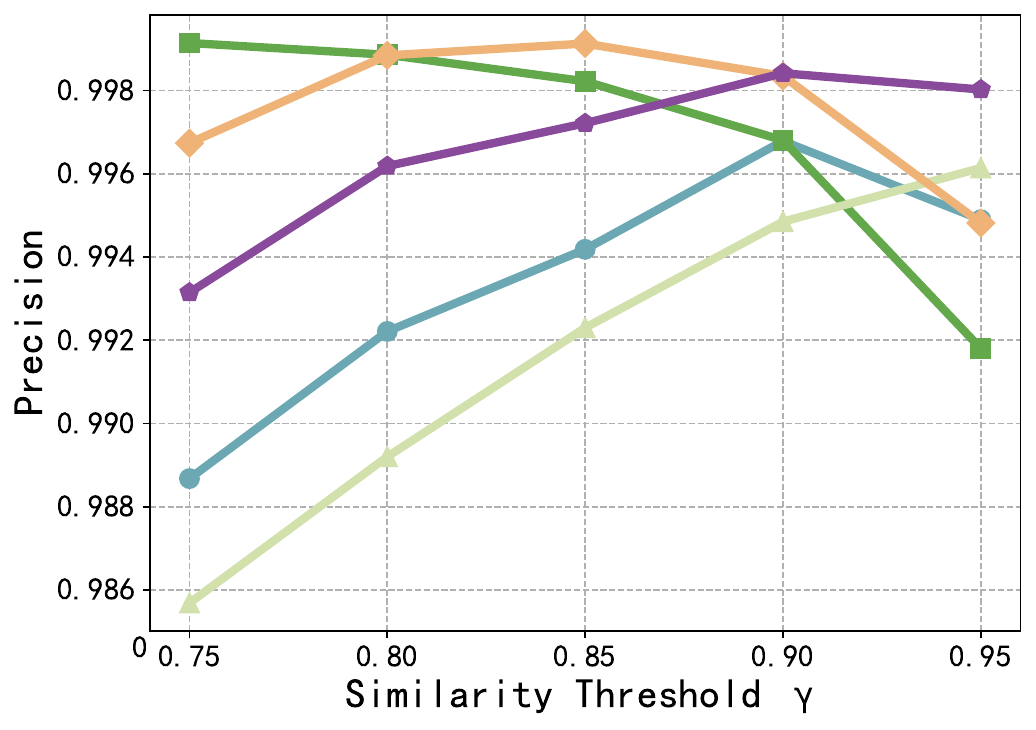}
        \caption{Precision vs $\gamma$}
        \label{fig:precision_vs_gamma}
    \end{subfigure}
    
    \vspace{0.5cm} 
    
    \begin{subfigure}[t]{0.48\linewidth}
        \centering
        \includegraphics[width=\linewidth]{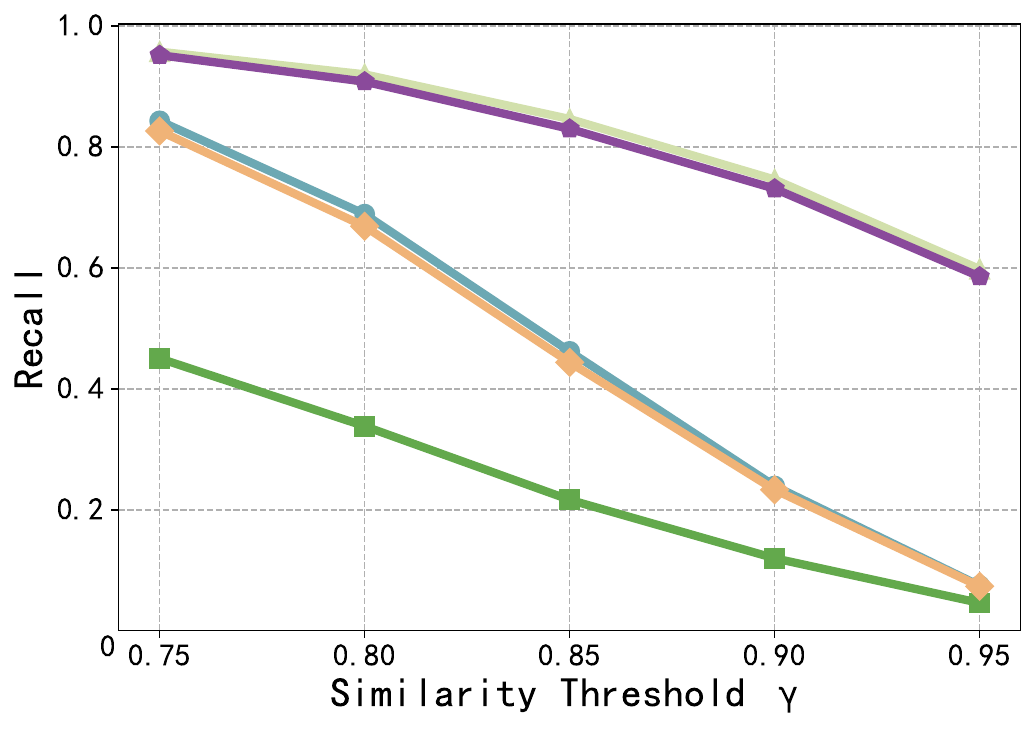}
        \caption{Recall vs $\gamma$}
        \label{fig:recall_vs_gamma}
    \end{subfigure}
    \hfill
    \begin{subfigure}[t]{0.48\linewidth}
        \centering
        \includegraphics[width=\linewidth]{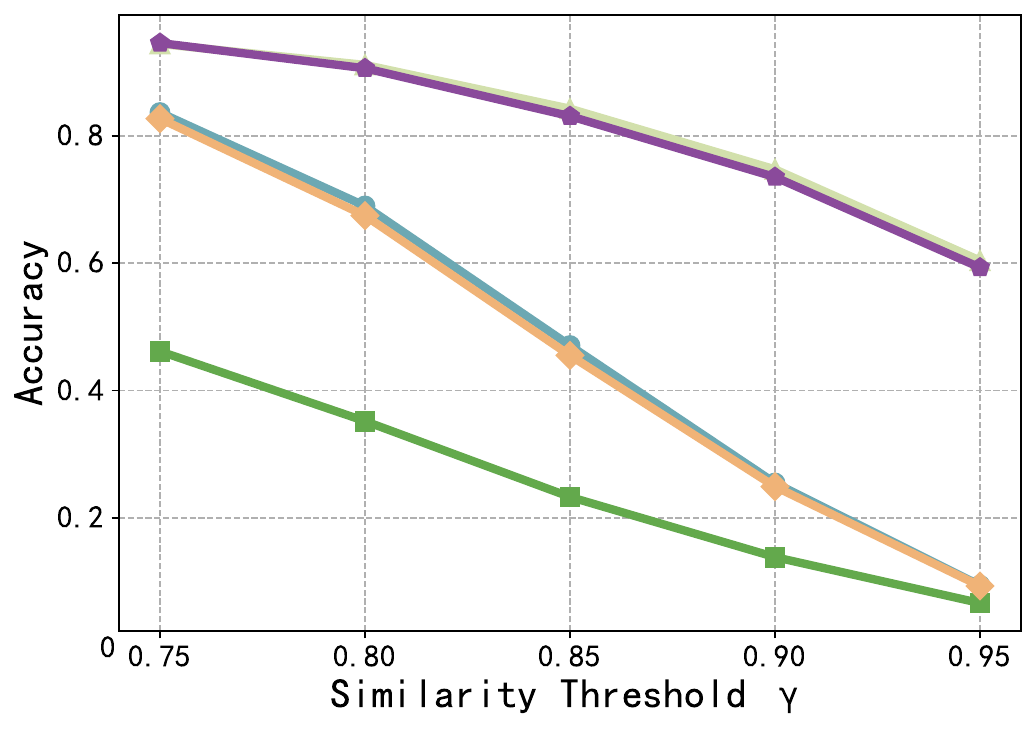}
        \caption{Accuracy vs $\gamma$}
        \label{fig:accuracy_vs_gamma}
    \end{subfigure}
    
    \vspace{0.3cm} 
    

    \caption{Effect of variation in similarity threshold ($\gamma$) on the performance of different methods. (a) F1 score vs $\gamma$, (b) Precision vs $\gamma$, (c) Recall vs $\gamma$, (d) Accuracy vs $\gamma$.}
    \label{fig:threshold_effects}
\end{figure}

Next, we adjust the similarity threshold $\gamma$, setting it to 0.75, 0.80, 0.85, 0.90, and 0.95, respectively, to observe the performance changes of different methods. Figure \ref{fig:threshold_effects} (a)-(d) shows the changes in F1 Score, Precision, Recall, and Accuracy for the five methods when the threshold changes. For F1 Score and Accuracy, AgentReuse and its variant OneIntent score higher. Overall, as the threshold increases, Recall and Accuracy gradually decrease, while Precision gradually increases. This is because when the threshold increases, the reuse mechanism becomes more conservative; only requests with extremely high similarity are judged as reusable, causing originally reusable requests to be judged as non-reusable, leading to a significant drop in Recall.

\begin{figure}[h]
    \centering
     \includegraphics[width=0.8\linewidth]{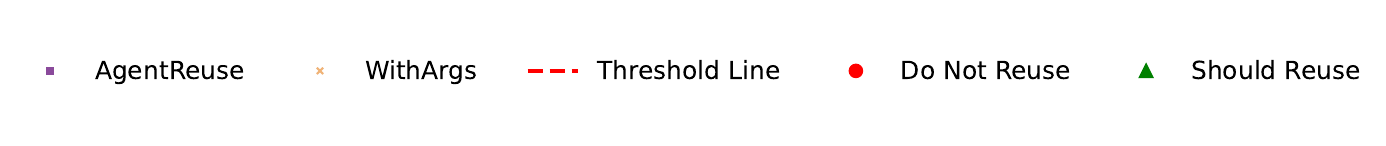}
    \includegraphics[width=0.95\linewidth]{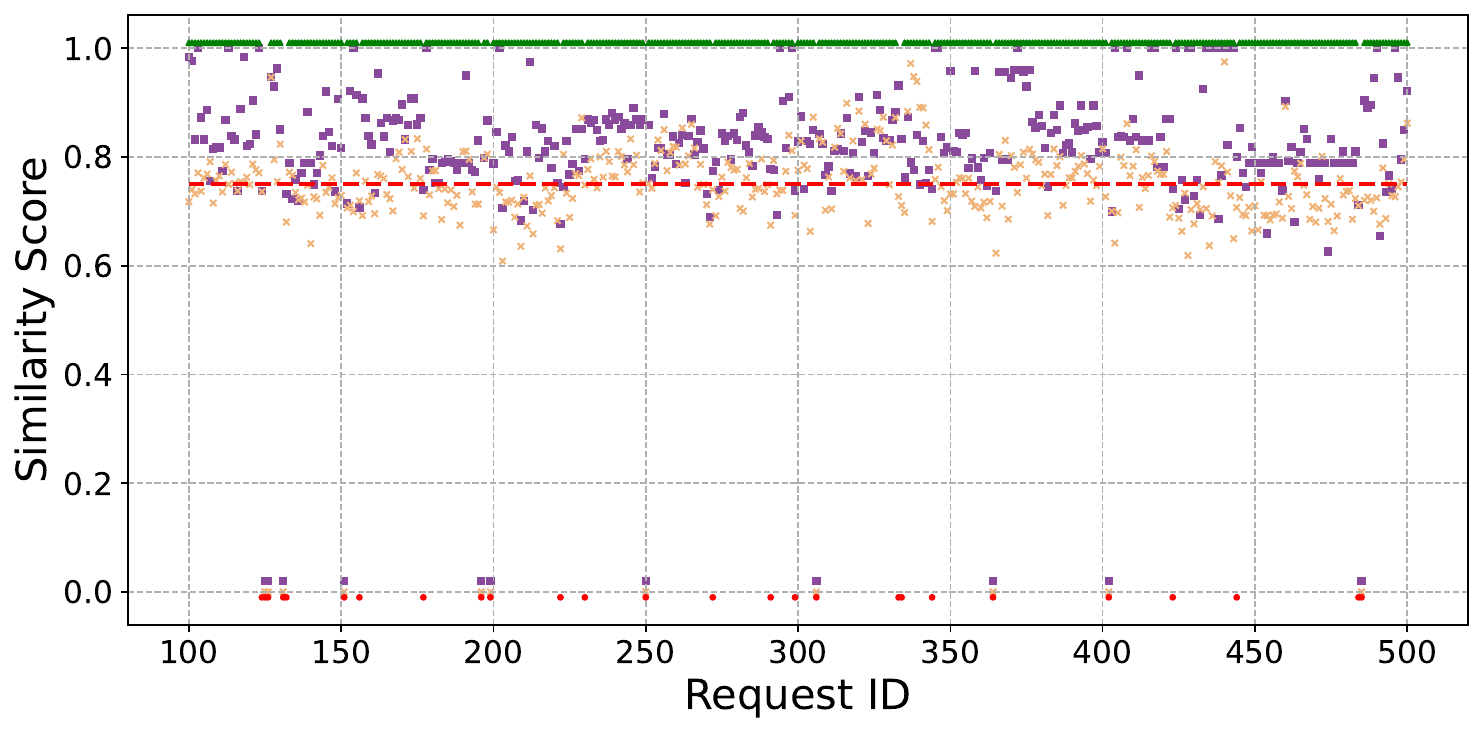}
    \caption{Comparison of similarity scores calculated by AgentReuse and WithArgs}
    \label{fig:sim_scores}
\end{figure}

Additionally, from Table \ref{tab:results} and Figure \ref{fig:threshold_effects}, we find that the WithArgs variant of AgentReuse performs poorly. WithArgs does not extract key parameters from the original request during similarity definition. Figure \ref{fig:sim_scores} shows the detailed similarity judgment comparison between AgentReuse and WithArgs for the same requests when the similarity threshold $\gamma = 0.75$. "Should Reuse" indicates that the request can factually reuse the plan generated by a previous request; "Should Not Reuse" indicates otherwise. For clarity, only results for Request IDs 100 to 500 are selected. As seen in Figure \ref{fig:sim_scores}, for the same request, because AgentReuse performs parameter extraction (removing dissimilar parts, i.e., key parameters, from similar sentences), AgentReuse obtains higher similarity scores, thereby achieving higher Recall (0.9513 vs 0.8262).

To analyze whether key parameter extraction is effective for structured plan reuse, we conducted experiments using AutoGen as the agent framework. The experimental standard was to execute two tests for the same user request: the first test used AgentReuse (extracting parameters, reusing the stored structured plan, executing it to get a response); the second test submitted the request directly to AutoGen to generate and execute a plan. Comparing the responses of the two tests, if the results were the same, parameter extraction and plan reuse were considered effective. We conducted experiments on 20 requests, repeating the experiment 5 times for each request (total 100 tests). In 93 tests, the responses were identical, indicating an effective reuse rate of 93\%. The 20 requests included 4 booking requests, 4 app launch requests, 2 query requests, 2 translation requests, 2 creation requests, 2 search requests, 2 download requests, 1 navigation request, and 1 undefined request.

This section analyzed the experimental results. The results show that when using the AgentReuse mechanism for similarity definition ($\gamma = 0.75$), the F1 Score reaches 0.9718 and Accuracy reaches 0.9459, both higher than comparison methods. Additionally, we analyzed the changes in metrics with varying thresholds and the effectiveness of removing key parameters. Finally, we analyzed the effectiveness of reuse.

\subsection{Extra Cost Analysis}
Compared to the execution process of existing LLM-driven agents, AgentReuse uses additional models, occupies extra storage space for caching, and correspondingly introduces extra latency. This section aims to demonstrate through experimental testing that the extra costs introduced by AgentReuse are acceptable.

\subsubsection{Storage Extra Cost Analysis}

1) \textbf{Model VRAM Usage}: The \texttt{bert-base-chinese} model used for intent classification and parameter identification has about 102.3 million parameters. Each parameter occupies 4B, theoretically occupying about 390.24 MB of VRAM. Experimental tests show the system allocated 444.00 MB, and actual usage was 391.16 MB. The \texttt{m3e-small} model used for vectorization has 24 million parameters, theoretically occupying about 91.55 MB. Tests show system allocation of 102.00 MB and actual usage of 92.12 MB. Although there is extra overhead during execution, the total VRAM usage remains at the MB level, which is acceptable for modern smartphones and other devices.

2) \textbf{Vectorized Text Usage}: Typically, vectorized text is stored as 32-bit floating-point numbers (4B each). \texttt{m3e-small} generates 512 dimensions, so each vectorized text occupies 2 KB. Including intent categories and FAISS index overhead (tested to be $<1$ MB), the extra memory cost for each request generally does not exceed 1 MB.

3) \textbf{Plan Text Usage}: AgentReuse stores plans in a structured format. Using UTF-8 encoding (Chinese characters occupy 3-4B, English/numbers 1B), the formatted plan described in Section \ref{sec:des} has 303 characters and occupies 705 bytes. On average, the extra storage overhead per plan text does not exceed 1 KB.

\subsubsection{Latency Extra Cost Analysis}

\begin{figure}[h]
    \centering
    \includegraphics[width=0.95\linewidth]{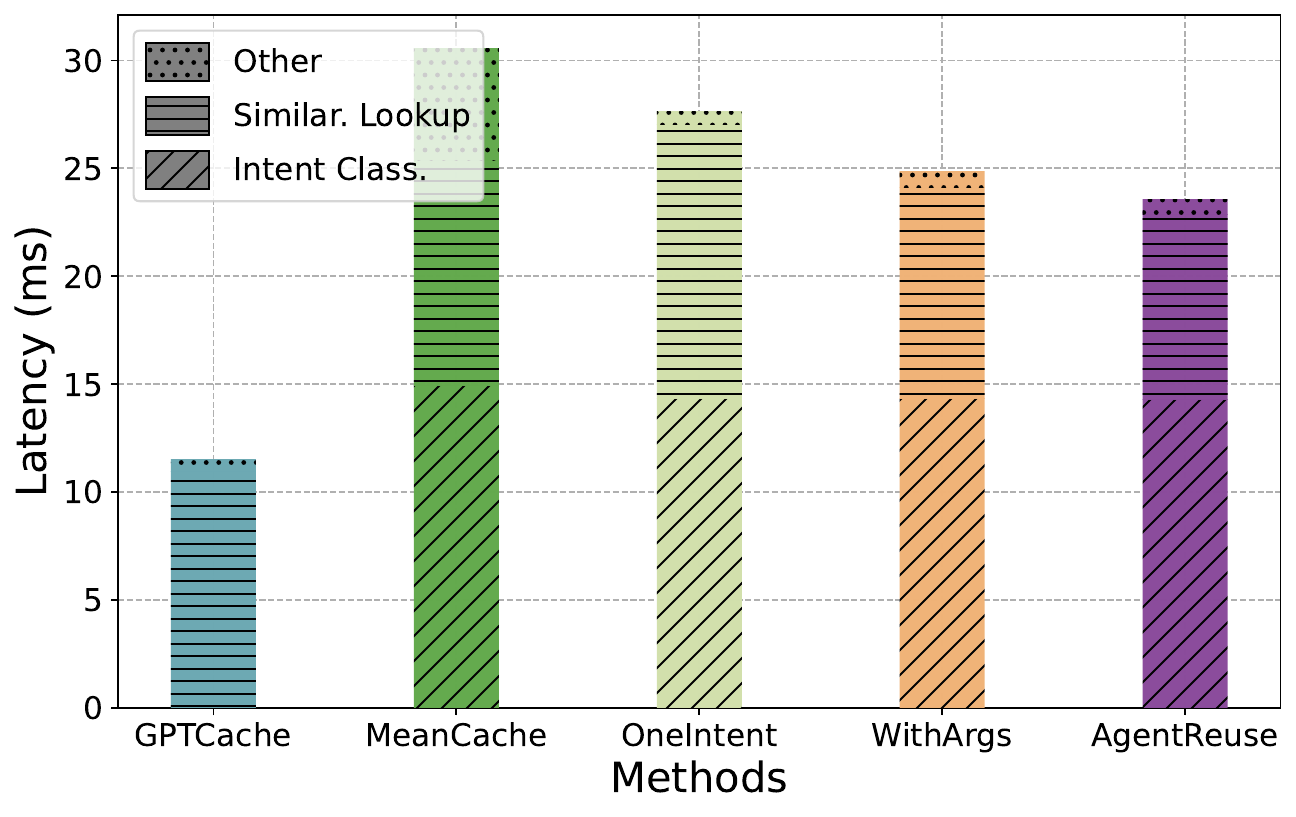}
    \caption{Composition of latency of different methods to process each request}
    \label{fig:latency_breakdown}
\end{figure}

For the reuse mechanisms of LLM-driven agents, the extra latency cost for MeanCache, OneIntent, WithArgs, and AgentReuse consists of intent classification, similarity search, and other latencies. GPTCache does not classify intent, so its latency consists of similarity search and other latencies. Other latencies mainly include text vectorization and storing content in the cache. Figure \ref{fig:latency_breakdown} shows the composition of latency costs per request for the 5 methods after 50 tests.

Figure \ref{fig:latency_breakdown} shows that GPTCache has the lowest total latency per request (11.528 ms) because it does not use an intent classification model, but its similarity definition performance is poor as mentioned in Section \ref{sec:subres}. Among methods using intent classification, AgentReuse has the lowest total latency (23.489 ms). It can be seen that AgentReuse and OneIntent have close performance in F1 Score and Accuracy, but AgentReuse has lower latency cost. OneIntent (a variant of AgentReuse) has 17.66\% higher latency than AgentReuse. The intent classification latency is similar, but OneIntent's similarity search latency is 47.94\% higher than AgentReuse because it searches all cached content, increasing the search space. Additionally, searching across all content leads to identifying some non-reusable requests as reusable, decreasing precision.

Furthermore, we observe that for the 4 methods excluding MeanCache, "other latencies" do not exceed 1 ms (all are 5.228 ms). MeanCache uses a PCA model, incurring extra latency for each request. for AgentReuse, the latency for formatting the plan and injecting parameters into the reusable plan was measured to be less than 1 ms on average, which is ignored in subsequent analysis.

In summary, the main extra costs of AgentReuse are approximately 500 MB of VRAM, less than 1 MB of memory for storing each vectorized text and structured plan, and 23.489 ms of latency per request. Considering that existing personal assistants already have intent classification models, AgentReuse would bring about 100 MB of extra VRAM cost and about 9.206 ms of processing latency per request. Reusing plans generated by agents not only saves plan generation latency but also reduces VRAM usage for running the LLM (typically >5 GB). Moreover, if using paid APIs like OpenAI, it saves costs. Therefore, the extra costs brought by AgentReuse are acceptable.

\subsection{Performance Gain Analysis}
This section analyzes the performance gains of using AgentReuse compared to using no reuse mechanism and using GPTCache. Since the reuse mechanism does not affect plan \textit{execution} latency, we only consider plan generation latency and the extra costs of the mechanism. We assume a plan generation latency of 31.8s and that all non-True Positive cases require plan generation.

In the experiment with AgentReuse ($\gamma = 0.75$), there were 2,464 requests that were True Positives (i.e., actually reusable and identified as such). For the 2,644 requests in the dataset, if all required LLM calls (no reuse), the latency would be 84,079.2s. If using AgentReuse (23.489 ms per request), processing 2,644 requests takes 62.1s. Generating plans for the 180 non-True Positive requests takes $180 \times 31.8s = 5,724s$. Thus, the total latency for AgentReuse is 5,786.1s. For GPTCache, there were 2,183 True Positives and 461 non-True Positives, resulting in a total latency of 14,690.28s. Therefore, compared to no reuse, AgentReuse reduces latency by 93.12\%. Compared to the LLM response caching method GPTCache, AgentReuse reduces latency by 60.61\%.

This section used the SMP dataset for experiments and analyzed the results. The results show that AgentReuse and its variant OneIntent outperform existing methods in similarity definition. Analysis of extra costs indicates that AgentReuse adds about 100 MB of VRAM, <1 MB of memory per request, and <10 ms latency compared to existing agent processes. Performance gain analysis shows that considering extra costs, AgentReuse reduces latency by 93.12\% compared to no reuse and 60.61\% compared to GPTCache.

\section{Discussion and Future Work}

\subsection{Multi-Intent Classification}
The main research object of this paper is the LLM-driven agent, specifically LLM-driven AIoT applications, such as Xiaomi's Xiao Ai and OPPO's Xiaobu Assistant. Traditionally, a typical user produces only one input to the agent, such as "Book a ticket from Hefei to Beijing for the day after tomorrow." Nowadays, with the development of society, smart cars are gradually entering households. Smart cars have more components, and at this time, a user's single sentence may contain multiple tasks, such as "Lower the front right window, raise the rear left window, turn on the wipers, and play music by David Tao." In this case, a multi-intent classification model needs to be adopted to classify intents and perform \ParamID in the text. After classification, the sentence is split into multiple parts, and then the AgentReuse mechanism is used for reuse in parallel. However, when there is a causal relationship between sentences, such as "If the indoor temperature exceeds 28 degrees Celsius, turn the air conditioner to cooling at 26 degrees Celsius," the sentence cannot be simply split into multiple parts. A new model needs to be introduced to process sentences with causal relationships, and then AgentReuse can be used to reuse previously generated plans. 

\subsection{Agent Runtime Environment under Serverless Computing Architecture}
Serverless computing, also known as server-imperceptible computing, is capable of achieving fine-grained resource management and rapid service scaling by executing tasks using containers \cite{jonas2019cloud, deng2022dependent}, and is regarded as one of the efficient computing paradigms for executing LLMs \cite{aws2024hugging, fu2024phantom}. Furthermore, as mentioned in Section \ref{sec:imple}, in LLM-driven agents, tool usage is also typically conducted within containers for system security considerations. Therefore, constructing an agent runtime environment under a serverless computing architecture is one of the future works of this paper. Regarding the reuse mechanism designed in this paper, converting unstructured texts into structured execution steps presents certain challenges when implementing effective plan reuse. When running agents under a serverless computing architecture, one can utilize system-level container execution tracking commands to trace input and output dependencies and execution orders among multiple containers executing tools. This allows for constructing more accurate formatted plan execution steps, thereby improving the effective plan reuse rate.


\section{Conclusion}
In today's era of artificial intelligence, agents are considered a promising path towards Artificial General Intelligence (AGI). This paper proposes and implements AgentReuse, a mechanism that can effectively reuse plans generated by LLM-driven agents, and tests the effectiveness and extra costs of this mechanism on a real-world dataset. Experimental results show that compared to existing methods, AgentReuse performs better in defining similarity between requests. By storing plan texts in a structured format, it achieves effective reuse of plans. Compared with not using a reuse mechanism, it can reduce latency by 93.12\%. In actual real-world operating environments, the tasks faced by agents are often more complex. As an exploratory work, the AgentReuse designed and implemented by us may not be able to handle all situations and needs to be modified and improved by combining characteristics from more practical scenarios. Today, LLMs have integrated into our lives, but LLM-driven agents have not yet been applied on a large scale. We hope this work can help promote the practical landing of agent applications, especially LLM-driven AIoT applications.

\clearpage
\bibliographystyle{plain}
\bibliography{references}

\end{document}